\documentclass{svjour3}                     
\smartqed  
\usepackage{graphicx}
\usepackage{bm}
\usepackage{amssymb,amsmath}
\usepackage{epstopdf}
\usepackage{epsfig}
\usepackage{color}
\usepackage{amsfonts}

\usepackage{amscd}%
\begin{document}

\title{ Exact Master Equation and Non-Markovian Decoherence
\\ for Quantum Dot Quantum Computing 
} 


\author{M. W. Y. Tu \and M. T. Lee \and W. M. Zhang$^*$}


\institute{Matisse Wei-Yuan Tu \at
              Department of Physics and Center for Quantum Information
              Science, National Cheng Kung University, Tainan, 70101 Taiwan. \\
            \emph{Present address:} Institute for Materials Science and Max
            Bergmann Center of Biomaterials, Dresden University of Technology,
            D-01069 Dresden, Germany.
           \and
           Ming-Tsung Lee \at
              National Center for Theoretical Science, Tainan, 70101 Taiwan. \\
            \emph{Present address:} Research Center for Applied Science, Academic Sinica,
            Taipei, 11529 Taiwan.  
           \and
           Wei-Min Zhang \at
              Department of Physics and Center for Quantum Information
              Science, National Cheng Kung University, Tainan, 70101 Taiwan;
              National Center for Theoretical Science, Tainan, 70101 Taiwan. \\
              $^*$\email{wzhang@mail.ncku.edu.tw}
}

\date{Received: date / Accepted: date}

\maketitle

\begin{abstract}
In this article, we report the recent progress on decoherence
dynamics of electrons in quantum dot quantum computing systems using
the exact master equation we derived recently based on the
Feynman-Vernon influence functional approach. The exact master
equation is valid for general nanostructure systems coupled to
multi-reservoirs with arbitrary spectral densities, temperatures and
biases. We take the double quantum dot charge qubit system as a
specific example, and discuss in details the decoherence dynamics of
the charge qubit under coherence controls. The decoherence dynamics
risen from the entanglement between the system and the environment
is mainly non-Markovian. We further discuss the decoherence of the
double-dot charge qubit induced by quantum point contact (QPC)
measurement where the master equation is re-derived using the
Keldysh non-equilibrium Green function technique due to the
non-linear coupling between the charge qubit and the QPC. The
non-Markovian decoherence dynamics in the measurement processes is
extensively discussed as well.


\end{abstract}

\section{Introduction}

The investigation of quantum coherence and quantum entanglement in
quantum dot systems has been attracting much attention in the
past decade because of its scalability for quantum computer. Due to
the rapid development of nanotechnology, implementing quantum
information processing using nanostructures is very prospective.
Among a diversity of nanostructures, a prototypical one is made by
fabrications on the interface of a heterostructure semiconductor
as a gate-defined region containing a set of coupled electronic
states. Electrodes are implanted around this region to control the
bias across it, and also several gates are implemented to adjust
the electronic states within the central area as well as their
couplings to the surrounding electrodes.  Because the electronic
structures of these systems can be well controlled, they have been
intensively and extensively studied for many purposes including
understanding fundamental physical issues and building useful
quantum devices. However, it is also because of its openness to the
controlling electrodes, the external gates and various degrees of freedom in
the host material, the central electronic region is intimately
entangled with the surrounding environment. Quantum entanglement among the
basic units for quantum information processing, i.e. quantum bits
or simply qubits, is the unique resource for exponentially speedup
over classical algorithms. However, quantum entanglement between
the qubit system and its environment is indeed a debt to quantum
information processing. The entanglement between the system and
its environment induces severe loss of quantum coherence of qubits
and destroys eventually the useful entanglement among qubits. It
has been realized that decoherence is the most difficult obstacle
for any successful quantum information processing. Therefore, the
investigation of decoherence dynamics due to the entanglement of
the qubit systems with the surrounding environment and the development
of decoherence control protocols become the central issue in the
study of quantum information processing. In this article, we will
present a general theory for the description of the real time
dynamics of a central electronic system coupled to
multi-electrodes. The theory is build on the exact master equation
for general nanostructures we derived recently using the
Feynman-Vernon influence functional approach \cite{TuPRB08}. This
exact master equation is capable to depict the detailed
decoherence dynamics of electronic states in the nanostructure
with arbitrary spectral density at an arbitrary initial
temperature of the reservoirs and an arbitrary bias applied to the
reservoirs.

Historically, since it was first proposed by Feynman and Vernon in
1963 \cite{fv63} for quantum Brownian motion (QBM) modelled as a
central harmonic oscillator linearly coupled to a set of harmonic
oscillators simulating the thermal bath, the influence functional
approach has been widely used to study dissipation dynamics in
quantum tunneling problems \cite{legt} and decoherence problems in
quantum measurement theory \cite{zuk01,zuk03}. In the early
applications, the master equation was derived for some particular
class of Ohmic (white-noise) environment \cite{cal}. The exact
master equation for the QBM with a general spectral density
(color-noise environments) at arbitrary temperature was obtained by
Hu and co-workers in 1992 \cite{hu}. Applications of the QBM exact
master equation cover various topics, such as quantum decoherence,
quantum-to-classical transition, quantum measurement theory, and
quantum gravity and quantum cosmology, etc.
\cite{Weiss99,Breuer02,Hu08}. Very recently, such an exact master
equation is further extended to the systems of two entangled
harmonic oscillators \cite{chou08} and two entangled optical fields
\cite{an07,an09} for the study of non-Markovian entanglement
dynamics. Nevertheless, using the influence functional approach to
obtain the exact master equation has been largely focused on the
bosonic environments in the past half century. The extension of the
influence functional approach to fermion environments is just
started \cite{TuPRB08}, where the Feynman path integral in terms of
fermion coherent states \cite{wmz90} must be used.

In fact, the investigation of the environmental effects on the
subsequent dynamics of the principal system one concerned is one of the
essential issues in many currently interesting research topics related to
quantum decoherence and dissipation phenomena. The quantum decoherence
and dissipation phenomena are also associated with
quantum non-equilibrium dynamics, a subject that has been
investigated over a half century yet is still not completely
understood. Quantum decoherence in quantum computations and quantum
information processes is one of the biggest challenges for practical
applications of quantum mechanics. Using the
Feynman-Vernon influence functional approach to understand
decoherence phenomena is fundamental in the sense that it allows to
completely integrate out the environmental degrees of freedom. As a
result, the back-action of the environment to the system can be
fully taken into account. The resulting non-perturbation (with
respect to the coupling between the system and its environment)
master equation is indeed desirable for quantum information
processing since the fast manipulations to qubit states requires a
strong coupling among the constitutes. It enables us to precisely explore the
real-time dynamics of the electron charge coherence in various different
nanostructures under different manipulating conditions.

On the other hand, quantum measurement (readout of qubit states)
also often induces severe decoherence. Therefore, precision
readout of qubit states is another big challenge in quantum
information processing. Again due to the rapid progress in
nanotechnology, electron state readout in nanostructures has been
investigated experimentally with qubit systems coupling to a
mesoscopic measurement device, such as single electron transistor
(SET) or quantum point contact (QPC). Not only its practical
application to quantum information processing, but also the
theoretical interests in the measurement-induced quantum
decoherence have attracted much attention. In particular, in the
investigation of quantum dot quantum computing with charge
qubits, the QPC has been served as an ultrasensitive electrometer.
The theoretical understanding of charge qubit measurement through
the QPC was previously treated based on the Markov approximation
\cite{Gurvitz,Goan01,mtlee06} in which the time scale of the qubit
dynamics is assumed much longer than that of the
tunneling-electron correlation in the QPC. The treatments based on
the Markov approximation, in fact, only describe the qubit
dynamics in the time-asymptotic quasi-equilibrium state. However,
for the quantum measurement in terms of SET or QPC, the qubit
decoherence occurs in the time scale of the same order of the
tunneling-electron correlation time in the measurement equipments,
where the non-Markovian dynamics of the qubit is
significant. In the second part of this article we shall discuss a fully
non-equilibrium theory we have developed recently
\cite{mtlee07,mtlee08} for understanding the qubit decoherence
induced by the QPC measurement, based on the Keldysh
non-equilibrium Green function technique \cite{keldysh}.

The Keldysh non-equilibrium Green function approach is well
developed for a perturbation treatment to the non-equilibrium
dynamics in many-body systems \cite{chou,Rammer}. This approach has
been used to study various transport phenomena in nanostructures
\cite{Jauho}. To make the non-equilibrium effect of the electrical
reservoir more transparent, a real-time diagrammatic technique was
constructed to diagrammatically calculate correlation functions of
the electrical reservoir order by order in the perturbation
expansion \cite{Schon94,Sch00}.  In the investigation of the
non-Markovian dynamics in the double dot charge qubit induced by the
QPC measurement, we developed an alternative real-time diagrammatic
technique \cite{mtlee07}.  The master equation for the charge qubit
dynamics is derived and expressed in terms of irreducible diagrams
up to all orders. The effect of the fluctuant measurement reservoir
due to the interaction with the qubit system can be fully taken into
account. The measurement-induced non-Markovian decoherence in the
qubit dynamics can be explicitly studied in this formulism.

\section{Exact master equation of electron transport systems}

The prototypical nanostructure under consideration can be generally
described by the following electronic Hamiltonian
\begin{align}
H=&\sum_{ij}\epsilon_{ij} d^\dag_i d_j +
\sum_{ij}U_{ij}d^\dag_i d_i d^\dag_j d_j \notag \\
&+\sum_{\alpha k} \epsilon_{\alpha k} c^\dag_{\alpha k}c_{\alpha
k} +\sum_{i \alpha k}(t_{i \alpha k}d^\dag_i c_{\alpha k} +t^*_{i
\alpha k} c^\dag_{\alpha k}d_i)~, \label{HH}
\end{align}
where the first two terms are the general Hamiltonian $H_D$ of
electrons in the central region of the nanostructure. For
simplicity, we shall ignore the electron-electron interaction in
this article, i.e., let $U_{ij}=0$. The third term is the
Hamiltonian $H_R$ describing the noninteracting electron reservoirs
(the source and drain electrodes, etc.) labeled by the index
$\alpha$. The last term is the electron tunneling Hamiltonian $H_T$
between the reservoirs and the central system.  $d^\dag_i (d_i)$ and
$c^\dag_{\alpha k} (c_{\alpha k})$ are the electron creation
(annihilation) operators of the central system and the surrounding
reservoirs, respectively. Also, throughout this work, we set
$\hbar=1$. The initial inverse temperature for the $\alpha$-lead is denoted
as $\beta_\alpha=1/k_{\rm B}T_\alpha$.

Using the Feynman-Vernon influence functional approach extended to
fermion coherent state representation \cite{wmz90}, we can integrate
out completely the degrees of freedom of the electron reservoirs and
obtain the following exact master equation \cite{TuPRB08},
\begin{align}
\dot{\rho}(t)=& -i[H_{D}'(t),\rho(t)] \nonumber\\& +\sum_{ij}\{
\Gamma_{ij}(t)[2d_{j}\rho(t)d^{\dag}_{i}
-d^{\dag}_{i}d_{j}\rho(t)-\rho(t)d^{\dag}_{i}d_{j}] \nonumber\\&
+\Gamma^{\beta}_{ij}(t)[d_{j}\rho(t)d^{\dag}_{i}-
d^{\dag}_{i}\rho(t)d_{j}-d^{\dag}_{i}d_{j}\rho(t)+\rho(t)d_{j}d^{\dag}_{i}]
\}~. \label{mst}
\end{align} This master equation determines completely
the quantum coherence dynamics of the central electronic system.
$\rho(t)$ is the reduced density matrix of the central system
at time $t$ after the environmental degrees of freedom are
completely eliminated. $
H_{D}'(t)=\sum_{ij}\epsilon'_{ij}(t)d^{\dag}_{i}d_{j} $ is the
corresponding renormalized Hamiltonian. Other two non-unitary terms
in the master equation describe the reservoir-induced dissipative
and noise processes with time-dependent dissipation and fluctuation
coefficients, $\Gamma(t)$
and $\Gamma^\beta(t)$. All these time-dependent coefficients in the
master equation are derived non-perturbatively and are exact. The
time-dependent fluctuations of the energy levels,
$\epsilon'_{ii}(t)$, and the transition couplings between states,
$\epsilon'_{ij}(t)$ for $i\ne j$, are the renormalization effects
that contain the energy level shifts and the coupling changes
between them, while $\Gamma(t)$ and $\Gamma^\beta(t)$ are the
dissipation and fluctuation effects that depict the full
non-Markovian decoherence dynamics. These are all the back-action
effects risen from the electron tunneling processes between the
central system and the reservoirs.

The explicit time dependencies of these coefficients in the master
equation are given as follows
\begin{subequations}
\label{td-coe}
\begin{align}
&\epsilon'_{ij}(t)= {i\over 2}[\dot{\bm u}(t)\bm u^{-1}(t)- (\bm
u^\dag)^{-1}(t)\dot{\bm u}^\dag(t)]_{ij},
\\
&\Gamma_{ij}(t)= -{1\over 2}[\dot{\bm u}(t)\bm u^{-1}(t)+ (\bm
u^\dag)^{-1}(t)\dot{\bm u}^\dag(t)]_{ij}, \label{gamma1}
\\
& \Gamma^\beta_{ij}(t)=[\dot{\bm u}(t){\bm u}^{-1}(t){\bm v}(t)
+{\bm v}(t)(\bm u^\dag)^{-1}(t)\dot{\bm u}^\dag(t)-\dot{\bm
v}(t)]_{ij}. \label{Gam-beta}
\end{align}
\end{subequations}
The elementary functions $\bm u (t)$ and $\bm v (t)$ are
time-dependent matrices following the dissipation-fluctuation
integral-differential equations \begin{subequations}\label{uv-eq}
\begin{align}
\label{ut-eq} \dot{\bm u}(\tau)+i\bm \epsilon {\bm u}(\tau)
+\sum_\alpha & \int_{t_0}^{\tau } d\tau' \bm F_\alpha (\tau-\tau')
{\bm
u}(\tau')=0, \\
\dot{\bm  v}(\tau)+i\bm \epsilon \bm  v(\tau) + \sum_\alpha &
\int_{t_0}^{\tau } d\tau' \bm F_\alpha (\tau-\tau') \bm v(\tau')
 =  \sum_\alpha \int_{t_0}^{t }d\tau'
      \bm F^\beta_\alpha (\tau-\tau')\bar{\bm u}(\tau'),
\label{vt-eq}
\end{align}
\end{subequations}
subjected to the boundary conditions $\bm u(t_{0})=\bm I$ and $\bm
v(t_{0})=0$, where $\bar{\bm u}(\tau)=\bm u^\dag(t-\tau+t_0)$. Introduce
the spectral density $J^{\alpha}_{ij}(\epsilon)=2 \pi \sum_{\mathbf{k}
\in\alpha} t^{*}_{i\alpha\mathbf{k}}t_{j\alpha\mathbf{k}}
\delta(\epsilon-\epsilon_{\alpha\mathbf{k}})$, then the integration
kernels in the above equations can be expressed as
$F_{ij}(\tau-\tau')=\sum_{\alpha}\int \frac{d\epsilon}{2\pi}
J^{\alpha}_{ij}(\epsilon)e^{-i\epsilon(\tau-\tau')}$, and
$F^{\beta}_{ij}(\tau-\tau')=\sum_{\alpha}\int \frac{d\epsilon}{2\pi}
f_{\alpha}(\epsilon)J^{\alpha}_{ij}(\epsilon)e^{-i\epsilon(\tau-\tau')}$,
where $f_{\alpha}(\epsilon)={1\over
e^{\beta_\alpha(\epsilon-\mu_{\alpha})}+1}$ is the initial fermi
distribution function of the reservoir $\alpha$ held at the chemical
potential $\mu_{\alpha}$.  These are the so-called
dissipation-fluctuation kernels and represent dissipation and
fluctuation effects on the central system due to its coupling to
electron reservoirs. The relation between these two kernels are then
seen from the dissipation-fluctuation theorem. The real time
non-equilibrium electron dynamics of the central systems is
fully described by the master equation.

\section{Coherence control of double dot charge qubit}

In the quantum computing scheme in terms of double quantum dots
where the electron charge degree of freedom is exploited, the
effects in deviating the coherency of charge dynamics are usually
summarized in the fluctuations of the inter-dot coupling and energy
splitting between the two local charge states. The amplitudes of
these fluctuations may be estimated from measurements of the noise
spectrum of electron currents and the minimum line width of elastic
current peak \cite{Fujisawa}. Parallel theoretical works have been
developed with different approaches in the literature for the
purposes of both simulating the experimental results and
understanding the physical mechanisms living in the double-dot. With
the master equation shown above, a complete investigation can be carried out
for the decoherence dynamics of the double dot charge qubit under coherence
controls. The decoherence of the electron
charges can be addressed with the back-reaction effects of the
reservoirs to the double dot (including not only the fluctuations of
the inter-dot coupling and energy splitting between the two local
charge states but also the non-unitary dynamical effect induced by
dissipations and noises) being fully taken into account.

\begin{figure}
  \includegraphics[width=0.4\textwidth]{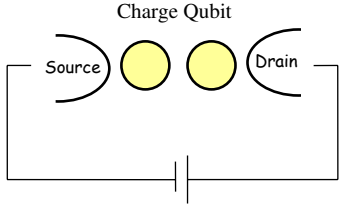}
\caption{(color online). A schematic plot of the double quantum
dot coupled to the source and drain electrodes.}
\label{fig:1}       
\end{figure}

For the double dot charge qubit (a schematic plot see
Fig.~\ref{fig:1}),
the corresponding Hamiltonian of the central system becomes
\begin{align} H_{\rm dot}= E_1 d^\dagger_1 d_1 + E_2 d^\dagger_2
d_2 +\Omega_0( d_{2}^{\dagger}d_{1}+d_{1}^{\dagger}d_{2})~, \label{HM-0a}
\end{align} where $E_{1}$ and $E_{2}$ are the on-site energies of
the left and the right dot respectively,  $\Omega_0$ is the tunnel
coupling between the two sites. The double quantum dot is coupled to
two electron leads on the left and the right side. Specifically, the
left dot is only coupled to the left lead while the right dot is
only coupled to the right lead.  To closely monitor the time
evolution of the electronic states in this system, it is more
convenient to specify the master equation in the particle number
basis of the no electron state $|0\rangle$, the state of one
electron in the left dot $|1\rangle$, one electron in the right dot
$|2\rangle$, and the double occupied state of two electrons in two
dots $|3\rangle$. Then the master equation is reduced to a set of
coupled rate equations in this basis. Furthermore, the double dot
charge qubit works in the single-electron regime, where the strong
Coulomb blockade (Coulomb repulsion between the two sites) plays an
important role.  In other words, the doubly occupied state
$|3\rangle$ must be excluded physically. This can be easily taken
into account by excluding the double occupancy from the rate
equations properly. The consequent rate equations excluding the
double occupancy read \cite{TuPRB08}
\begin{subequations}
\label{i-re}
\begin{align}
\dot{\rho}_{00}&=\tilde{\Gamma}_{11}\rho_{11} +
\tilde{\Gamma}_{21}\rho_{12} + \tilde{\Gamma}_{12}\rho_{21} +
\tilde{\Gamma}_{22}\rho_{22}+{\rm Tr}\Gamma^{\beta}\rho_{00}, \\
\dot{\rho}_{11}&=-\tilde{\Gamma}_{11}\rho_{11}
+\tilde{\Xi}^*_{-}\rho_{12} +\tilde{\Xi}_{-}\rho_{21}
-\Gamma^\beta_{11}\rho_{00},
 \\ \dot{\rho}_{22}&=-\tilde{\Gamma}_{22}\rho_{22}
+\tilde{\Xi}^*_{+}\rho_{12} +\tilde{\Xi}_{+}\rho_{21}
-\Gamma^\beta_{22}\rho_{00}, \\
\dot{\rho}_{12}&=[-i\epsilon' -{1\over 2}{\rm
tr}\tilde{\Gamma}]\rho_{12} +\tilde{\Xi}_+\rho_{11}
+\tilde{\Xi}_-\rho_{22} -\Gamma^\beta_{12}\rho_{00},
\end{align}
\end{subequations}
where $\tilde{\Gamma}(t)=2\Gamma(t)+\Gamma^{\beta}(t)$,
$\tilde{\Xi}_\pm(t)=\pm i\epsilon'_{12}(t)-{1\over2} \tilde{\Gamma}_{12}(t)$
and $\epsilon'(t)=\epsilon'_{11}(t)-\epsilon'_{22}(t)$, $\Gamma(t)$ and $\Gamma^\beta(t)$ are
the time-dependent coefficient matrices in the master equation (\ref{mst})
with $i=1,2$. We can reproduce other rate equations from
(\ref{i-re}) in the corresponding Born-Markov limit used in the
literature \cite{Stoof96,Gurvitz96,Brandes}.

To study the decoherence dynamics of the charge qubit, the
spectral density must be specified. Here we use a Lorentzian
spectral density. For the double dot charge qubit system, since
the left (right) dot only couples to the left (right) electrode,
the spectral density can be written explicitly as
\begin{align}
J^{\alpha}_{ij}(\epsilon)={\Gamma_{\alpha}W^2_{\alpha}\over(\epsilon
-E_{i})^{2}+W_{\alpha}^{2}}\delta_{ij} ~,~~
\end{align}
with $\alpha=L (R)$ for $i=1 (2)$.  The Lorentzian spectral widths
$W_{L,R}$ are indeed the bandwidths of the densities of states for
the source and drain, respectively. They depict the characteristic
times of the corresponding reservoirs.  The larger the widths
$W_{L,R}$ are, the shorter the correlation time of the reservoirs
give and the lesser the memory effect shows. In the wide band limit
(WBL) that is widely used in the literature, the spectral density
becomes energy independent. As a matter of fact, all the memory
effect is wiped out. The coupling constants $\Gamma_{L,R}$ are the
tunneling rates of electron tunneled into the double dot from the
source and the drain, respectively. The tunneling rates
$\Gamma_{L,R}$ describe the strength of the tunneling process
between the dot and the reservoirs covering the leakage effect, and
are experimentally adjustable through gate voltages. Varying the
coupling strength to the environment and the correlation time scale
of the reservoirs, we are able to analyze realistically the
decoherence dynamics of the double dot charge qubit.

\begin{figure}
  \includegraphics[width=0.5\textwidth]{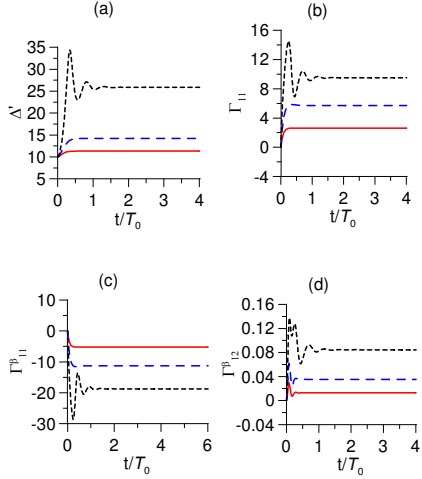}
\caption{(color online). The time dependence of transport
coefficients in the master equation where $\Delta'(t)=2\epsilon'_{12}(t)$.
The inter-dot tunnel coupling
$\Omega_{0}$ is set to be $5\mu$eV with the fixed chemical potentials
at $\mu-E=50\mu$eV and the Lorentzian spectral width $W=20\mu$eV .
The red line, the blue dashed line and the black dot lines
correspond to $\Gamma=5, 10$ and $25 \mu$eV, respectively.}
\label{fig:2}       
\end{figure}

For manipulating the charge qubit coherence, the device is set at
the resonant condition $E_{1}=E_{2}=E$.  The fermi surfaces of the
two electrodes are aligned $\mu_{L}=\mu_{R}=\mu$ above the resonant
level $\mu-E>0$ so that electrons are kept in the dot with minor
probability of leakage. To simplify the problem, we consider only
the symmetric double dot system where $\Gamma_{L}=\Gamma_{R}=\Gamma$
and $W_{L}=W_{R}=W$. The electron temperature is kept at $100mK$
throughout the rest of the calculations. The time-dependence of the
transport coefficients in the master equation is essential for the
non-Martkovian dynamics during the coherence controls of charge
qubit states. Here we first examine the time dependent behavior of
these transport coefficients. Fig.~\ref{fig:2} shows an example of
the time dependencies of those coefficients at a few different
strengths of coupling (tunneling rate) between the qubit and
electron reservoirs, where the coupling strength can be controlled
by tuning the gate voltages in experiments. The time is in unit of
$T_0=2\pi/\sqrt{(E_1-E_2)^2+4\Omega_0^2} =\pi/\Omega_0$. As one can see the
stronger the coupling between the qubit and the environment is, the
larger fluctuation the coefficients shows that reveals the
non-Markov feature.
As it is known the rate between the bandwidth $W$ and the tunneling
rate $\Gamma$ tells how important the non-Markovian dynamics can be.
The width $W$ characterizes the correlation time of the electron
reservoirs, a large width $W$ corresponds to a short characteristic
time of the reservoir. Therefore, a smaller $W/\Gamma$ will make the
non-Markovian process more obvious in electron dynamics. This
feature is shown by the time-dependence of the transport
coefficients, as we can see from Fig.~\ref{fig:2}. When $W/\Gamma$
becomes large enough, the time-dependence of the transport
coefficients will disappear and the Markovian limit may be reached.
Equivalently speaking, a larger width $W$ corresponds to a shorter
correlation time of the electron reservoir, the coefficients at
larger width reach their steady limit sooner.

\begin{figure}
  \includegraphics[width=0.5\textwidth]{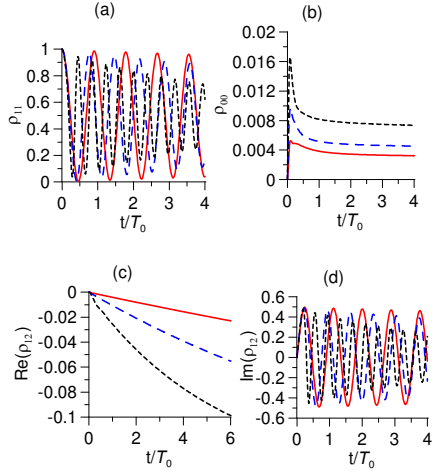}
\caption{(color online). The time evolution of the reduced density
matrix elements for the charge qubit solved from the master
equation. All the parameters are the same as that given in
Fig.~\ref{fig:2}}
\label{fig:3}       
\end{figure}

Correspondingly, the dynamics of the charge qubit is also examined
at different coupling strengths by plotting the time evolution of
the reduced density matrix elements. Fig.~\ref{fig:3}(a)-(d) show
the time evolution of the electron population occupying the first
dot given by $\rho_{11}(t)$, the probability of empty dots, i.e. the
leakage effect by $\rho_{00}(t)$, and the real and imaginary parts of
the off-diagonal matrix element $\rho_{12}(t)$.
For the symmetric dots ($E_1=E_2$) concerned here, the initial
energy eigenstates of the double dots are given by the bonding and
anti-bonding states, $|\pm \rangle = \frac{1}{\sqrt{2}}(|1\rangle
\pm |2\rangle)$. The relaxation and decoherence of the qubit states
are then determined by the decays of the matrix elements $\langle +
|\rho|+\rangle =\frac{1}{2}(\rho_{11}+\rho_{22}) + {\rm
Re}\rho_{12}$ and $\langle
+|\rho|-\rangle=\frac{1}{2}(\rho_{11}-\rho_{22}) -i {\rm
Im}\rho_{12}$, respectively. In other words, the decay rate of ${\rm
Re}\rho_{12}(t)$ is the relaxation time $T_1$ (because
$\rho_{11}+\rho_{22} \simeq 1$) while the decay rates of
$\rho_{11}(t), \rho_{22}(t)$ or Im$\rho_{12}(t)$ give the
decoherence time $T_2$, as we have discussed in \cite{TuPRB08}.
These properties are clearly shown in Fig.~\ref{fig:3}, where the
oscillation decays in Fig.~\ref{fig:3}(a) and (d) depict the
decoherence dynamics while Fig.~\ref{fig:3}(c) describes the
relaxation process (energy dissipation) and Fig.~\ref{fig:3}(b)
shows the small leakage effect. We find that the decay of the charge
coherent oscillation is well described by a simple exponential decay
for the off-diagonal reduced density matrix elements. The diagonal
matrix elements (populations) are better described by a
sub-exponential law when charge leakage is not negligible, otherwise
simple exponential decay is better for all the cases. The relaxation
time $T_1$ and the decoherence time $T_2$ can be extracted from the
exact numerical solution $\mbox{Re}\rho_{12}(t)$ and
$\mbox{Im}\rho_{12}(t)$, respectively, with the result $T_1 \leq
2T_2$ being of the order of a few nanoseconds or less for a broad
parameter range we used \cite{TuPRB08}.

\begin{figure}
  \includegraphics[width=0.5\textwidth]{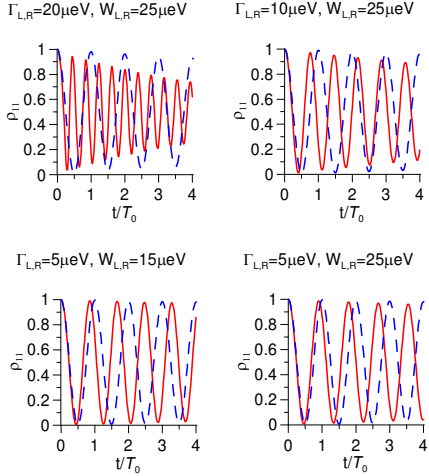}
\caption{(color online). A comparison of the exact solution (red
lines) with the Markov limit (blue dashed lines), where the
inter-dot tunnel coupling $\Omega_{0}=5\mu$eV with the fixed chemical
potential at $\mu-E=50\mu$eV.}
\label{fig:4}       
\end{figure}

Furthermore, a stronger tunneling coupling $\Gamma$ to the
electron reservoirs (compare to the Lorentzian spectral width $W$)
leads to a faster decay of the coherent charge qubit oscillation
and also causes severer shifts in the charge oscillation
frequency.  The non-Markovian decoherence dynamics of charge qubit
is dominated by two major effects, the memory effect and the
leakage effect in the double-dot gated by electrode reservoirs.
The former becomes a dominate effect when the time scale (the
inverse of the width $W$) of the reservoirs is comparable to the
time scale ($\sim T_0$) of the double-dot. The latter becomes
important when the electron tunneling strength between the
reservoirs and dots is tuned to be large.  Strengthening the
couplings between the reservoirs and dots disturbs the charge
coherence in the double-dot significantly. However, reasonably
raising up the chemical potentials $\mu_{L,R}$ can suppress charge
leakage and maintain charge coherence. The left uncontrollable
decoherence factor is the spectral width which characters how many
electron states in the reservoirs effectively involving in the
tunneling processes between the dots and the reservoirs. It is
quite interest to see that in any case, the Markov limit
(based on the rate equations derived by Brandes and Vorrath
under Born-Markov approximation but without including the phonon
coupling \cite{Brandes}) deviates largely from the exact
solutions, as shown in Fig.~\ref{fig:4}. It indicates that
quantum information processing is indeed non-Markovian with the
double dot charge qubit.

\section{Charge qubit with QPC measurement}

\begin{figure}
  \includegraphics[width=0.4\textwidth]{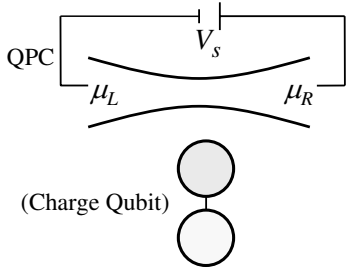}
\caption{A schematic plot for the QPC measurement of double dot
charge qubit.}
\label{fig:5}       
\end{figure}

In this section, we present a study of the charge qubit
measurement using QPC in the tunnel junction regime. The
transmissions of all tunneling channels cross the QPC barrier are
required to be small enough in this regime such that the electron
tunneling becomes sensitive to the qubit state. The qubit
information can then be extracted from the output signal of the
QPC, while the back-action of the measurement to the qubit states
can also be taken into account. In order to explore the qubit
decoherence induced by the QPC measurement, the corresponding
Hamiltonian takes the form \cite{Gurvitz}
\begin{align}
H=H_{\rm dot} +\sum_{\alpha,k}\epsilon_{\alpha k}a_{\alpha
k}^{\dag}a_{\alpha k}+
\sum_{kk'}(q_{kk'}a_{Rk'}^{\dag}a_{Lk}+q^\dag_{kk'}a_{Lk}^{\dag}a_{Rk'}).
\label{eq:coupling}
\end{align}
where $H_{\rm dot}$ is the Hamiltonian of the double dot charge
qubit given by Eq.~(\ref{HM-0a}). The second term is the
Hamiltonian of the QPC electronic reservoirs consisting of the
source $(\alpha=L)$ and the drain $(\alpha=R)$ electrodes with the
energy levels $\epsilon_{\alpha k}$, and the chemical potentials
$\mu_{L}$ and $\mu _{R}$, respectively. The last term is the
interaction Hamiltonian describing the electron tunneling
processes through the QPC with a qubit-state dependent hopping
amplitude $q_{kk'}=\Omega_{kk'}-\delta
\Omega_{kk'}d_{1}^{\dag}d_{1}$. The single electron occupies on
the first dot leads to a variation in the barrier of the QPC, and
the electron hoping amplitude is thus modified from $\Omega_{kk'}$
to $\Omega'_{kk'}=\Omega_{kk'}-\delta \Omega_{kk'}$. The charge
qubit state is measured through the electron tunneling across the
source and the drain. A schematic device is plotted in
Fig.~\ref{fig:5}.

\begin{figure}
  \includegraphics[width=0.5\textwidth]{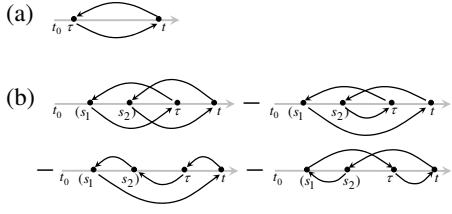}
\caption{Irreducible diagrams: (a) and (b) are the leading $K_{\rm
ir}^{(0)}(t-\tau )$ and the second leading-order $ K_{\rm
ir}^{(2)}(t-\tau )$ contributions, respectively.} \label{fig:6}
\end{figure}

To explore the measurement-induced decoherence of charge qubit,
the QPC can be considered as an environment coupled to the charge
qubit. Then the qubit dynamics is determined by the master
equation for the reduced density operator $\rho(t)=$tr$_{R}[\rho
_{\rm tot}(t)]$, where $\rho_{\rm tot}(t)$ is the total density
operator for the whole system of the qubit plus QPC reservoir, and
the partial trace tr$_{R}$ integrates over all the degrees of
freedom of the QPC reservoir. Since the qubit-reservoir coupling
involves two-body interactions, the influence functional approach
for integrating over all the degrees of freedom of the QPC
reservoir cannot be exactly carried out. Therefore, we developed
a real-time diagrammatic technique based on the Keldysh
non-equilibrium Green function approach to study the decoherence
effect of non-equilibrium electrodes on the qubit. The master
equation for the charge qubit coupled with the QPC measurement
has been derived in terms of the irreducible diagrams to all
the orders in the perturbation expansion \cite{mtlee07,mtlee08},
\begin{equation}
\dot{\rho}_(t)=-i\big[ H_{S},\rho(t)\big] -\sum_{n=0}^\infty
\int_{t_{0}}^{t}d\tau K^{(2n)}_{\rm ir}(t-\tau )*\rho(\tau ),
\label{eq:eomirr}
\end{equation}
where the kernel expansion $K_{\rm ir}^{(2n)}(t-\tau )*\rho(\tau
)$ consists of all the irreducible diagrams containing
$2n+2$-particle correlations. The irreducible diagrams are defined
by all \textit{connected} topology-independent diagrams along the
real time axis. The connected diagram means that each loop in the
diagram should intersect with other loops at least once.
Physically, in the tunneling processes classified by $K_{\rm
ir}^{(2n)}(t-\tau )$, there must exist at least two particles
correlating together in arbitrary time interval of the time period
from $t_{0}$ to $t$.  The leading and the second-order irreducible
diagrams are shown in Fig.~\ref{fig:6}, where each vertex depends
on the qubit operator via $q_{kk'}$ and the connecting free
propagator specifies the non-equilibrium effect of the tunneling
electron QPC on the qubit through the Keldysh Green function
matrix. For the detailed diagrammatic rules and the classification
of topology-independent irreducible diagrams, please refer to
\cite{mtlee07,mtlee08}.

Since the QPC is often used for continuous measurement of charge
qubit states, the coupling between the qubit and the QPC can be
treated as a weak coupling. For the study of this QPC
measurement-induced non-Markovian decoherence to the qubit states,
it may be sufficient to consider only the leading order
contribution to the master equation. The master equation (up to
the leading order) is given as follows
\begin{eqnarray}
&&\dot{\rho}(t)=-i\big[H_{S},\rho(t)\big] -\int_{t_0}^{t}d\tau
\big[R_0,[[K(t-\tau), \tilde{\rho}(\tau) ]]\big], \label{eq:kernj}
\end{eqnarray}
where $\tilde{\rho}(\tau ) \equiv e^{-iH_{S}(t-\tau )}\rho(\tau
)e^{iH_{S}(t-\tau )}$, the double bracket $\left[ \left[
A,B\right] \right] \equiv AB-(AB)^{+} $, the operator $R_0 = R(0)$
and $R(t)$ is defined as $R(t) =\cos \theta (\left| e\right\rangle
\left\langle e\right| -\left| g\right\rangle \left\langle g\right|
)-\sin \theta (e^{i\gamma t}\left| g\right\rangle \left\langle
e\right| +e^{-i\gamma t}\left| e\right\rangle \left\langle
g\right| )$ with $\gamma =\sqrt{4\Omega_0^{2}+(E_{1}-E_{2})^{2}}$ the
energy difference between the ground state $\left| g\right\rangle
$ and the excited state $\left| e\right\rangle $ of the qubit and
$\theta =\cos ^{-1}[(E_{1}-E_{2})/\gamma ]$. The operator
$K(t)=k(t)R(t)$, and $k(t)$ the reservoir correlation function
that characterizes the QPC structure. It can be expressed
explicitly as
\begin{align}
k(t)=&\int \frac{d\epsilon d\epsilon'}{(2\pi)^{2}}e^{i(\epsilon
-\epsilon')t}\big\{f_{L}(\epsilon)[1-f_{R}(\epsilon')]J(\epsilon,
\epsilon')+f_{R}(\epsilon )[1-f_{L}(\epsilon')] J(\epsilon',
\epsilon) \big\}, \label{eq:k}
\end{align}
where $f_{L,R}(\epsilon )=1/(1+\exp \beta (\epsilon -\mu _{L,R}))$
are the Fermi-Dirac distribution functions for the source and the
drain, $J(\epsilon ,\epsilon')$ is the spectral density of the QPC
reservoir which is defined by
\begin{align}
J(\epsilon ,\epsilon')=&\pi^{2} g_L(\epsilon) g_R(\epsilon')
   |\delta \Omega(\epsilon,\epsilon')|^2 ~,  \label{spectral}
\end{align}
with $g_{L,R}(\epsilon )$ being the density of states of the
source and the drain, and $\delta \Omega(\epsilon,\epsilon')$ the
difference of the energy-dependent hopping amplitude (between the
source and the drain) without and with the occupation of the first
dot.

\begin{figure}
  \includegraphics[width=0.3\textwidth,angle=-90]{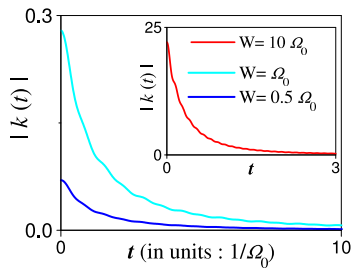}
\caption{(color online). The amplitude of the reservoir time
correlation function during the measurement with different width
$W$. Here we take the parameters $w=2\Omega_0$, $\beta =1/\Omega_0$,
$V_{d}=20\Omega_0$ and $\Gamma_{d} = \Omega_0^{2}/4$, respectively. The
amplitude is plotted in units $\Omega_0^{2}$.} \label{fig:7}
\end{figure}

A close connection between the qubit decoherence and the
tunneling-electron fluctuation is revealed through the reservoir
time correlation function $k(t-\tau)$ of Eq.~(\ref{eq:k}). The
reservoir time correlation function describes the variation of the
tunneling-electron correlation during the measurement. As we
mentioned in the introduction, the previous investigations for the
QPC measurement-induced decoherence of charge qubit are mainly
focused on the Markov limit \cite{Gurvitz,Goan01,mtlee06} in which
the time scale of the qubit dynamics is assumed much longer than
that of the tunneling-electron correlation in the QPC. Such
theoretical treatments based on the Markov approximation
describe only the qubit dynamics in the time-asymptotic
quasi-stationary state. Mesoscopically, the qubit decoherence
occurs in the time scale of the same order of the
tunneling-electron correlation in the QPC, where the non-Markovian
dynamics of the qubit is significant \cite{mtlee08}. To study the
qubit decoherence in the non-Markovian regime, a Lorentzian-type
spectral density for the QPC reservoir is considered here
\begin{equation}
J(\epsilon,\epsilon')=\frac{ \Gamma_d W^2
}{(|\epsilon-\epsilon'|+ w)^{2}+W^{2}}, \label{eq:resolutf}
\end{equation}
where the constant $\Gamma_{d}$ specifies the decay rate of the
qubit state due to the interaction of the qubit electron with the QPC, and the factor
$|\delta \Omega|^{2}$ in Eq.~(\ref{spectral}) has been absorbed into
$\Gamma_{d}$. The spectral width $W$ determines the correlation time
scales of tunneling electrons across the QPC barrier, and $w$ is an
additional parameter to characterize the variation of the QPC internal
structure due to the interaction with the qubit electron. In the
literature, the wide band limit, $W \gg \gamma$ [so that
$J(\epsilon,\epsilon') \rightarrow \Gamma_{d}$], has been often
used to specify the QPC reservoir structure which simply leads to
the Markov limit at the time-asymptotic
quasi-stationary state \cite{Gurvitz,Goan01,mtlee06}.  In
Fig.~\ref{fig:7}, the reservoir time correlation function is
plotted. The corresponding correlation time scale is determined by
the half width of the profile. The wider the half width of
$|k(t)|$ is, the longer correlation time scale the
tunneling-electron fluctuation maintains. The correlation time
scale increases with $W$ decreasing.

\begin{figure}
  \includegraphics[width=0.45\textwidth,angle=-90]{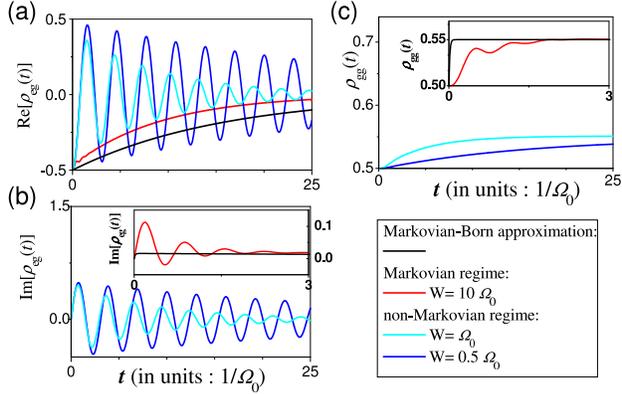}
\caption{(color online). The charge qubit dynamics for different width $W$.
The charge qubit is set initially in the $ |1\rangle $ state. The other
parameters are taken as the same as in Fig.~\ref{fig:7}. (a) and
(b) for the charge qubit dephasing, and (c) the charge qubit
relaxation.} \label{fig:8}
\end{figure}

To make the qubit decoherence feature apparent due to the QPC
measurement, we concentrate again on the charge qubit with symmetric
coupled dots $E_{1}=E_{2}$. Thus the energy scale of qubit
dynamics is simply characterized by $\gamma =2\Omega_0$ ($\theta
=\pi/2$). The master motion for the reduced density matrix
becomes,
\begin{subequations}
\begin{align}
\dot{\rho}_{eg}(t)=&-i\gamma \rho _{eg}(t)-2\int_{t_{0}}^{t}d\tau
|k(t-\tau )|\cos [\phi (t-\tau )]\big\{\rho _{eg}(\tau )-\rho
_{ge}(\tau )\big\}, \label{eq:eg}
\\
\dot{\rho}_{gg}(t)=&2\int_{t_{0}}^{t}d\tau |k(t-\tau )|\big\{\cos
[\phi^-_\gamma(t-\tau )]\rho _{ee}(\tau ) - \cos[\phi^+_\gamma
(t-\tau)]\rho_{gg}(\tau )\big\}, \label{eq:gg}
\end{align}
\end{subequations}
where the matrix elements are defines as $\rho
_{ij}(t)=\left\langle i|\rho(t)|j\right\rangle $ with $\left|
i,j\right\rangle $ being the ground state or the excited state of
the qubit, $\phi (t)$ is a time-dependent phase of the reservoir
correlation function $k(t-\tau)=|k(t-\tau)|e^{-i\phi (t-\tau)}$,
and we have also defined $\phi^\pm_\gamma \equiv \phi(t-\tau) \pm
(t-\tau)\gamma$. The qubit dynamics with different $W$ are
simulated in Fig. \ref{fig:8}.
As we can see, with a very large width $W (=10\Omega_0)$ the charge
qubit undergoes severe decoherence, where the time correlation of
tunneling-electron fluctuation is smeared, and the reservoir memory
effect on the qubit dynamics is washed out. In other words, the
qubit decoherence dynamics is Markov in this regime, which almost
coincides with the result of the Born-Markov approximation obtained
in the literature \cite{Gurvitz,Goan01,mtlee06}, see the black
curves in Fig.~\ref{fig:8}.  However, if the QPC structure can be
recasted such that the width $W$ is comparable to the energy scale
of the charge qubit, then non-Markovian processes becomes dominate
in the qubit dynamics. Tunneling electrons in this case propagate
with a larger correlation time. The charge qubit simply undergoes a
oscillation, see Fig.~\ref{fig:8}. It is very interesting to see
that the qubit decoherence is indeed suppressed in this case. On the
other hand, when a large bias voltage is applied, it usually
shortens the correlation time of the tunneling-electron fluctuation.
A large amount of electrons tunneling across the QPC barrier at a
large bias leads to a sharp profile for the time correlation
function $|k(t)|$ such that the qubit dynamics evolves into a severe
decoherence. However, it is interesting to see from Fig.~\ref{fig:8}
that even though a large bias voltage ($V_d=20 \Omega_0$) is
applied, the qubit decoherence can still be suppressed if the QPC is
fabricated to reach a narrower profile of the spectral density. The
qubit states can then be read out by a significant current signal
without much decoherence. In conclusion, more useful non-Markovian
dynamics can be manifested when the explicit effect of a realistic
spectral density with a finite correlation time scale is taken into
account, where the qubit decoherence can be suppressed during the
QPC measurement.

\section{Prospection}
In this article we have made an extensive discussion on the
non-Markovian decoherence dynamics of a double dot charge qubit
induced from the coherence controls and the QPC measurement, using
the master equation we obtained from the Feynman-Vernon influence
function approach and the Keldysh non-equilibrium Green function
technique. For quantum information processing, decoherence dynamics
of individual qubit is no doubt a key concern in these electronic
systems. On the other hand, decoherence dynamics of entanglement
between two qubits \cite{Yu04,Bello07} is another important issue
for quantum information processing. Some progress on decoherence
entanglement dynamics in cavity optical systems has been reported
based on exact master equations \cite{chou08,an07,an09}. Further
applications to the decoherence dynamics of electron charge and spin
entanglement in terms of the quantum dots \cite{Petta,Zhang07} are
expected. The full non-equilibrium electron dynamics and the real
time monitoring of spin polarization processes in quantum dots are
future topics. These together with other physical properties in
various nanostructures, such as Kondo effect and Fano resonance,
etc., as well as the transient dynamics of electronic quantum
transports are under progress. In short, the theory we discussed in
this work can be used to study not only the problem of decoherence
but also many other interesting physical phenomena in various
nanostructures.


\begin{thebibliography}{}
%
%

\bibitem{TuPRB08} M. W. Y. Tu and W. M. Zhang, \textit{Non-Markovian
decoherence theory for a double-dot charge qubit}, Phys. Rev. B {\bf
78}, 235311 (2008).

\bibitem{fv63} R. P. Feynman and F.L. Vernon, \textit{The theory of
a general quantum system interacting with a linear dissipative
system}, Ann. Phys. {\bf 24}, 118 (1963).

\bibitem{legt} A. J. Legget, S. Chakravarty, A. T. Dorsey, M. P. A. Fisher,
A. Garg and W. Zwerger, \textit{Dynamics of the dissipative
two-state system}, Rev. Mod. Phys. \textbf{59}, 1 (1987).

\bibitem{zuk01} W. H. Zurek, \textit{Decoherence and the transition from
quantum to classical}, Phys. Today {\bf 44}  (10), 36 (1991);

\bibitem{zuk03} W. H. Zurek, \textit{Decoherence, einselection, and the
quantum origins of the classical}, Rev. Mod. Phys. \textbf{75}, 715 (2003).

\bibitem{cal} A. O. Caldeira and A. J. Leggett, \textit{Path integral
approach to quantum Brownian motion}, Physica A \textbf{121}, 587
(1983).


\bibitem{hu} B. L. Hu, J. P. Paz, and Y. H. Zhang, \textit{Quantum Brownian
motion in a general environment: Exact master equation with
nonlocal dissipation and colored noise}, Phys. Rev. D \textbf{45},
2843 (1992).

\bibitem{Weiss99} U. Weiss, \textit{Quantum dissipative systems}, (World Scientific,
Singapore, 1999).

\bibitem{Breuer02} H. P. Breuer and F. Petruccione, \textit{The theory of
open quantum systems}, Oxford University Press, Oxford, (2002).

\bibitem{Hu08} E. Calzetta and B. L. Hu, \textit{Nonequilibrium quantum
field theory}, Cambridge University Press, New York, (2008).

\bibitem{chou08}C. H. Chou, T. Yu, and B. L. Hu,  \textit{Exact Master
Equation and Quantum Decoherence of Two Coupled Harmonic
Oscillators in a General Environment}, Phys. Rev. E \textbf{77},
011112 (2008).

\bibitem{an07} J. H. An and W. M. Zhang, \textit{Non-Markovian entanglement
dynamics of noisy continuous-variable quantum channels}, Phys.
Rev. A \textbf{76}, 042127 (2007).

\bibitem{an09} J. H. An, M. Feng and W. M. Zhang, \textit{Non-Markovian
decoherence dynamics of entangled coherent states}, Quantum Infom.
Comput. \textbf{9}, 0317 (2009).

\bibitem{wmz90} W. M. Zhang, D. H. Feng, and R. Gilmore, \textit{Coherent
states: Theory and some applications}, Rev. Mod. Phys. {\bf 62}, 867
(1990).

\bibitem{Gurvitz} S. A. Gurvitz, \textit{Measurements with a noninvasive
detector and dephasing mechanism}, Phys. Rev. B \textbf{56},
15215-15223 (1997).

\bibitem{Goan01} H. S. Goan, G. J. Milburm, H. M. Wiseman, and H. B. Sun,
\textit{Continuous quantum measurement of two coupled quantum dots
using a point contact: A quantum trajectory approach}, Phys. Rev. B
\textbf{63}, 125326 (2001).


\bibitem{mtlee06} M. T. Lee and M. W. Zhang, \textit{Decoherence induced by
electron accumulation in a quantum measurement of charge qubits},
Phys. Rev. B, \textbf{74}, 085325 (2006).

\bibitem{mtlee07}  M. T. Lee and W. M. Zhang, \textit{Non-equilibrium theory
of charge qubit decoherence in the quantum point contact
measurement}, arXiv: 0708.2581 (2007).

\bibitem{mtlee08} M. T. Lee and W. M. Zhang, \textit{Non-Markovian suppression of
charge qubit decoherence in the quantum point contact measurement},
J. Chem. Phys. \textbf{129}, 224106 (2008).


\bibitem{keldysh} L. V. Keldysh, \textit{Diagram Technique for Nonequilibrium
Processes}, Zh. Eksp. Teor. Fiz. \textbf{47}, 1515-1527 (1964)[Sov.
Phys. JETP, \textbf{20}, 1018-1026 (1965)].

\bibitem{chou} K. C. Chou, Z. B. Su, B. L. Hao, and L. Yu, \textit{Equilibrium
and nonequilibrium formalisms made unified}, Phys. Rep.
\textbf{118}, 1 (1985).

\bibitem{Rammer} J. Rammer and H. Smith, \textit{Quantum field-theoretical
methods in transport theory of metals}, Rev. Mod. Phys. \textbf{58},
323-359 (1986).

\bibitem{Jauho} H. Haug and A. P. Jauho, \textit{Quantum Kinetics in
Transport and Optics of Semiconductors}, Springer, Berlin, (1996).

\bibitem{Schon94} H. Schoeller and G. Sch\"{o}n, \textit{Mesoscopic quantum transport:
Resonant tunneling in the presence of a strong Coulomb interaction},
Phys. Rev. B \textbf{50}, 18436-18452 (1994).

\bibitem{Sch00} H. Schoeller and J. K\"{o}nig, \textit{Real-Time
Renormalization Group and Charge Fluctuations in Quantum Dots},
Phys. Rev. Lett. \textbf{84}, 3686-3689 (2000).

\bibitem{Fujisawa} T. Fujisawa, T. Hayashi, and S. Sasaki, \textit{Time-dependent
single-electron transport through quantum dots}, Rep. Prog. Phys.
\textbf{69}, 759 (2006).

\bibitem{Stoof96} T. H. Stoof and Yu. V. Nazarov, \textit{Time-dependent resonant
tunneling via two discrete states}, Phys. Rev. B \textbf{53}, 1050
(1996).

\bibitem{Gurvitz96} S. A. Gurvitz and Ya. S. Prager, \textit{Microscopic derivation
of rate equations for quantum transport}, Phys. Rev. B \textbf{53}, 15932 (1996).

\bibitem{Brandes} T. Brandes and T. Vorrath, \textit{Adiabatic transfer of electrons
in coupled quantum dots}, Phys. Rev. B \textbf{66}, 075341 (2002).

\bibitem{Yu04} T. Yu and J. H. Eberly, \textit{Finite-time disentanglement
via spontaneous emission}, Phys. Rev. Lett., \textbf{93}, 140404
(2004).

\bibitem{Bello07} B. Bellomo, R. Lo Franco, and G. Compagno,
\textit{Non-Markovian Effects on the Dynamics of Entanglement},
Phys. Rev. Lett., \textbf{99}, 160502 (2007).

\bibitem{Petta} J. R. Petta, A. C. Johnson, J. M. Taylor, E. A. Laird, A.
Yacoby, M. D. Lukin, C. M. Marcus, M. P. Hanson, and A. C.
Gossard, \textit{Coherent manipulation of coupled electron spins
in semiconductor quantum dots}, Science \textbf{309}, 2180 (2005).

\bibitem{Zhang07} W. M. Zhang, Y. Z. Wu, C. Soo, and M. Feng,
\textit{Charge-to-Spin conversion of electron entangled states and
spin-interaction-free solid-state quantum computation}, Phys. Rev.
B 76, 165311 (2007).

\end{thebibliography}
\end{document}